\RequirePackage{fix-cm}
\documentclass[smallextended]{svjour3}       
\smartqed  
\usepackage{graphicx}
\newcommand\etal {{\it et al.}}

\newcommand\al{\alpha}

\newcommand\ga{\gamma}
\newcommand\de{\delta}

\newcommand\ka{\kappa}
\newcommand\la{\lambda}

\newcommand\si{\sigma}

\newcommand\ps{\psi}

\newcommand\Ga{\Gamma}

\newcommand\Si{\Sigma}

\newcommand\mn{{\mu\nu}}

\newcommand\fr[2]{{{#1} \over {#2}}}
\newcommand\half{{\textstyle{1\over 2}}}

\newcommand\fracn[2]{{\textstyle{{#1}\over {#2}}}}

\newcommand\lsim{\mathrel{\rlap{\lower4pt\hbox{\hskip1pt$\sim$}}
    \raise1pt\hbox{$<$}}}
\newcommand\gsim{\mathrel{\rlap{\lower4pt\hbox{\hskip1pt$\sim$}}
    \raise1pt\hbox{$>$}}}
\newcommand\sqr[2]{{\vcenter{\vbox{\hrule height.#2pt
         \hbox{\vrule width.#2pt height#1pt \kern#1pt
         \vrule width.#2pt}
         \hrule height.#2pt}}}}

\newcommand\thpr{{these proceedings}}

\newcommand\pt[1]{\phantom{#1}}
\newcommand\ol[1]{\overline{#1}}

\newcommand\vb[2]{e_{#1}^{{\pt{#1}}#2}}
\newcommand\ivb[2]{e^{#1}_{{\pt{#1}}#2}}
\newcommand\uvb[2]{e^{#1#2}}

\newcommand\ab{\overline{a}{}}

\newcommand\cb{\overline{c}{}}

\newcommand\sbar{\overline{s}{}}

\newcommand\mt{m^{\rm T}}
\newcommand\ms{m^{\rm S}}

\newcommand\afb{(\ab_{\rm{eff}})}

\newcommand\afbx[1]{(\ab^{#1}_{\rm{eff}})}
\newcommand\cbx[1]{(\cb^{#1})}

\newcommand\afbe{\afbx{e}}

\newcommand\cbw{\cbx{w}}
\newcommand\afbw{\afbx{w}}

\newcommand\lrpartial{\raise 1pt\hbox{$\stackrel\leftrightarrow\partial$}}

\newcommand\lrDmu{\stackrel{\leftrightarrow}{D_\mu}}

\newcommand\atext{$a_\mu$}
\newcommand\btext{$b_\mu$}
\newcommand\ctext{$c_{\mu\nu}$}
\newcommand\dtext{$d_{\mu\nu}$}
\newcommand\etext{$e_\mu$}
\newcommand\ftext{$f_\mu$}
\newcommand\gtext{$g_{\la\mu\nu}$}
\newcommand\Htext{$H_{\mu\nu}$}

\newcommand\G{G_N}

\newcommand{\beq}{\begin{equation}}
\newcommand{\eeq}{\end{equation}}
\newcommand{\bea}{\begin{eqnarray}}
\newcommand{\eea}{\end{eqnarray}}
\newcommand{\bit}{\begin{itemize}}
\newcommand{\eit}{\end{itemize}}

\begin{document}

\title{Antimatter-Gravity Couplings, and Lorentz Symmetry
}


\author{Jay D.\ Tasson 
}


\institute{
Proceedings of the 11th International Conference on Low Energy Antiproton Physics (LEAP2013)
held in Uppsala, Sweden, 10-15 June 2013 \\
\line(1,0){110}\\
Jay D.\ Tasson \at
              Physics and Astronomy Department, Carleton College, Northfield, MN 55901, USA\\
              Present Address:\\
              Physics Department, St.\ Olaf College, Northfield, MN 55901, USA\\
              \email{tasson1@stolaf.edu}           
}

\date{}

\maketitle

\begin{abstract}
Implications of possible CPT and Lorentz violation
for antimatter-gravity experiments
as well as other antimatter tests
are considered in the context of the
general field-theory-based framework
of the Standard-Model Extension (SME).
\end{abstract}

\section{Introduction}
\label{intro}

The realm of antimatter
is an area in which many predictions of our current best theories of physics,
the Standard Model of particle physics and General Relativity
remain unverified.
Thus antimatter experiments
provide the opportunity
to place our existing theories
on a stronger experimental foundation.
Lorentz symmetry,
along with the associated CPT symmetry \cite{cpt},
is an aspect of both of our existing theories
that can be tested in antimatter experiments.

Beyond improving the experimental foundation
of the Standard Model and General Relativity,
antimatter experiments are among those
that offer the potential to detect new physics at the Planck scale
through searches for Lorentz and CPT violation
with low-energy physics \cite{jtrev}.
Standard lore holds that our current theories
are the low-energy limit of a more fundamental theory at the Planck scale.
Candidates for the fundamental theory
include suggestions such as string theory.
The possibility of CPT and Lorentz violation
has been shown to arise in some candidates for the underlying theory
including string theory \cite{ksp}
and other scenarios \cite{mav},
hence searches for violations
of these symmetries offers the opportunity to detect Planck-scale physics
with existing technology \cite{ksp}.

The SME offers a comprehensive field-theoretic framework
for investigating Lorentz and CPT symmetry
as an expansion about known physics \cite{ck,akgrav,nonmin}.
As such,
it contains known physics and offers the power to predict the outcome of relevant experiments.
It should be emphasized that the SME is not a specific model,
but a comprehensive test framework
ideally suited for a broad search.
Presently,
no compelling evidence for CPT or Lorentz violation has been found.
As such,
a broad and systematic search 
may offer a more efficient way of seeking such violations
than the consideration of many specific models
having unclear relationships to each other.
From this perspective,
a few models that illustrate aspects of the general framework
are useful;
however, the time for more aggressive model building
will come when new physics is found.
This proceedings contribution
reviews SME-based projects and proposals in the area of antimatter
as well as an SME-based model that illustrates several possibilities
in antimatter-gravity experiments.

\section{Basics}
\label{theory}

The QED-extension limit of the gravitationally coupled SME \cite{akgrav}
provides the basic theory relevant for the discussion to follow.
The action for the theory can be 
schematically presented as
\beq
S = S_\ps + S_{\rm gravity} + S_A.
\label{action}
\eeq
Here the first term
is the gravitationally coupled fermion sector,
the pure-gravity sector is provided by the second term,
and the final term is the photon sector.
Each of the terms above
consists of known physics
along with all Lorentz-violating terms that can be constructed
from the associated fields.

In this work
we specialize to the popular minimal-SME limit,
involving operators of dimension 3 and 4.
The minimal fermion-sector action can be 
presented explicitly as follows:
\beq
S_\ps = 
\int d^4 x (\half i e \ivb \mu a \ol \ps \Ga^a \lrDmu \ps 
- e \ol \ps M \ps),
\label{fermion}
\eeq
where
\bea
\Ga^a
&\equiv & 
\ga^a - c_{\mu\nu} \uvb \nu a \ivb \mu b \ga^b
- d_{\mu\nu} \uvb \nu a \ivb \mu b \ga_5 \ga^b
\nonumber\\
&&
- e_\mu \uvb \mu a 
- i f_\mu \uvb \mu a \ga_5 
- \half g_{\la\mu\nu} \uvb \nu a \ivb \la b \ivb \mu c \si^{bc}, \\
M
&\equiv &
m + a_\mu \ivb \mu a \ga^a 
+ b_\mu \ivb \mu a \ga_5 \ga^a 
+ \half H_{\mu\nu} \ivb \mu a \ivb \nu b \si^{ab},
\label{mdef}
\eea
and \atext, \btext, \ctext, \dtext, \etext, \ftext, \gtext, \Htext\ 
are coefficient fields for Lorentz violation.
Gravitational couplings
occur here via the vierbein $\vb \mu a$
and the covariant derivative.
The Minkowski-spacetime fermion-sector action
can be recovered in the limit $\vb \mu a \rightarrow \de^a_\mu$.

Leading contributions to the pure-gravity sector action
take the form
\beq
S_{\rm gravity} = \fr {1}{2\ka} \int d^4x e (R - u R 
+s^\mn R^T_\mn + t^{\ka\la\mu\nu} C_{\ka\la\mu\nu}),
\label{grav}
\eeq
where $R^T_\mn$ is the traceless Ricci tensor,
and $C_{\ka\la\mu\nu}$ is the Weyl tensor.
The first term provides the Einstein-Hilbert action.
The coefficient field $s^\mn$ in the third term
is responsible for
the relevant Lorentz-violating signals 
in the post-newtonian analysis \cite{lvpn}.
The coefficient field $t^{\ka\la\mu\nu}$
provides no contributions
in the post-Newtonian analysis,
while the coefficient field $u$ is not Lorentz violating.
The photon sector,
containing Maxwell electrodynamics
along with associated Lorentz violation,
is provided by the partial action $S_A$.
Since it is not directly relevant for the discussion to follow,
the explicit form of $S_A$ is omitted here,
though in general it is of considerable interest
and has been the subject of a large number of tests \cite{tables}.

The nature of the coefficient fields appearing above
can be understood in two basic ways:
explicit Lorentz violation and spontaneous Lorentz violation.
In the context of explicit Lorentz violation,
the content of the coefficient fields is imposed as an external choice,
where as in the context of spontaneous Lorentz violation,
the coefficient fields are dynamical fields within the theory
that receive vacuum expectation values via 
the spontaneous breaking of Lorentz symmetry.
Here spontaneous Lorentz violation
can be thought of as analogous to the spontaneous breaking of $SU(2) \times U(1)$
symmetry in the Standard Model.
Unlike electroweak-symmetry breaking,
the vacuum values that arise from spontaneous Lorentz violation
are vector or tensor objects
known as coefficients for Lorentz violation,
which can be thought of as establishing preferred directions in spacetime.
For Minkowski-spacetime experiments seeking CPT- and Lorentz-violating effects
associated with the vacuum values,
the distinction between explicit and spontaneous Lorentz-symmetry breaking
is not relevant,
and spacetime-independent coefficients for Lorentz violation are typically assumed,
both because they could be thought of as a leading term 
in an expansion of a more general function,
and because energy and momentum are conserved in this limit.

The Minkowski-spacetime limit
provides the relevant theory
for the nongravitational tests 
considered in Sec.\ \ref{ngt}.
In this limit,
the pure-gravity sector 
is irrelevant, gravitational couplings
in the fermion-sector \ref{fermion}
are neglected,
and constant coefficients provide the relevant contribution
to the coefficient fields.
Theoretical tools for the analysis of a variety of experiments
have been obtained from the general theory
including the associated relativistic quantum mechanics \cite{bkr},
the corresponding nonrelativistic Hamiltonian \cite{kl},
and the associated classical Lagrangian \cite{kr}.
Antimatter applications of these Minkowski-spacetime results
are considered in Sec.\ \ref{ngt}.

In the context of gravitational studies,
it has been shown that explicit Lorentz violation
is typically incompatible with the Riemann geometry 
of existing gravity theories \cite{akgrav},
though more general geometries may admit explicit breaking \cite{akfin}.
Hence progress in gravitational studies
requires consideration of spontaneous breaking.
Here spacetime-independent vacuum values
are still considered,
but geometric consistency requires one to additionally consider
the effects of certain contributions to the fluctuations about the vacuum values.
Sec.\ \ref{gt} considers gravitational couplings.
Lorentz-violating effects
in a gravitational context
can stem from the pure-gravity sector
or gravitational couplings in the other sectors of the theory.
Reference \cite{lvpn} provides numerous detailed experimental proposals for investigating
the coefficient (vacuum value) $\sbar^\mn$ 
associated with the coefficient field $s^\mn$.

An analysis of gravitational couplings in the fermion sector
is provided by Ref.\ \cite{lvgap}.
This includes the implications of the fluctuations in the coefficient fields
that must be addressed in a gravitational analysis
along with some additional theoretical issues.
A detailed analysis of the implications of spin-independent coefficient fields
\atext, \ctext,\ and \etext\
is then provided for a large class of experiments.
Vacuum values associated with these fields are denoted $\afb_\mu$,
for the countershaded combination \cite{akjt}
associated with the \atext\ and \etext\ fields,
and $\cb_\mn$ for the vacuum value associated with \ctext. 
The vacuum values introduced here correspond to the coefficients for Lorentz violation
discussed in Minkowski spacetime.
The experimental implications associated with these fermion-sector coefficients,
including those relevant for antimatter,
are summarized in Sec.\ \ref{gt}.

\section{Nongravitational CPT Tests with Antimatter}
\label{ngt}

Over 1000 experimental and observational measurements have been performed
in the context of the SME \cite{tables}.
Some methods associated with these measurements
are based on CPT- and Lorentz-violating modifications to the 
energy levels of atoms \cite{clock,hhbar}.
As the Earth revolves around the Sun
and rotates on its axis,
the orientation and boost of an experiment in the lab change.
In the context of studies of Lorentz violation,
such changes in boost and orientation imply annual and sidereal variations of energy levels.
Since Lorentz-violating effects on various levels typically differ,
Lorentz violation can be investigated
by searching for relative changes in levels with time \cite{clock}.
A subset of the SME coefficients for Lorentz violation
are also CPT violating,
and the comparison of the spectrum of hydrogen
with that of antihydrogen
would also provide sensitivity to associated coefficients \cite{hhbar}.
An experiment aiming to obtain such sensitivity
via a measurement of the hyperfine Zeeman line
is presently in preparation \cite{bjhyp}.

Lorentz- and CPT-violating effects
that are difficult to detect in matter experiments
may in some cases be more readily seen
in comparisons of matter and antimatter.
The isotropic `invisible' model (IIM) \cite{lvgap,akifc}
was constructed as a special limit of the SME
to illustrate a scenario in which Lorentz violation
would produce a significantly larger effect in 
tests involving antimatter
than in tests with only matter.
The model is constructed by first noting
that in any give inertial frame $O$,
a subset of Lorentz-violating operators
in the SME preserve rotational symmetry.
Isotropic Lorentz violation is generated in that frame by
setting the coefficients of the other operators to zero.
In another frame $O^\prime$ that is boosted with respect to $O$,
rotation invariance will be violated.  
The IIM then assumes
that $(b^p)_{T^\prime}$ and isotropic $(d^p)_{\Xi^\prime\Xi^\prime}$
are the only nonzero coefficients for Lorentz violation
in the CMB frame,
and it further assumes
$(b^p)_{T^\prime} = k m^p (d^p)_{T^\prime T^\prime}$,
for a suitable choice of constant $k$.
Nonzero coefficients
$(b^p)_J$ and $(d^p)_{JT}$
are then generated in the Sun-centered frame,
which is boosted relative to the CMB frame.
The dominant signals
in terrestrial experiments with hydrogen
are in the hyperfine structure
and involve
$(b^p)_J - m^p (d^p)_{JT}$,
which vanishes for suitable $k$.
Thus in the IIM,
hydrogen experiments can only detect
effects suppressed by at least one power
of the boost of the Earth around the Sun,
which is about $10^{-4}$
and requires experiments sensitive to annual variation.
However,
dominant effects in experiments with antihydrogen
involve the combination
$(b^p)_J + m^p (d^p)_{JT}$,
since \btext\ is a CPT odd coefficient, while \dtext\ is CPT even.
This implies unsuppressed signals would occur in the hyperfine structure
of antihydrogen in the IIM,
making it a field-theoretic toy model
in which the effects of CPT and Lorentz violation
would be at least $10^4$ times greater in antihydrogen
than those in hydrogen
or other nonrelativistic neutral matter.

The shifting of atomic energy levels
discussed above constitute one method of searching for Lorentz and CPT violation
in antimatter experiments.
The comparison of energy levels for trapped particles
with those of trapped antiparticles
also leads to CPT-violation sensitivities in the SME \cite{bkr}.
This method has led to experimental results
for coefficients for Lorentz violation associated with the 
electron \cite{etrap}
and proton \cite{ggtrap}
and ongoing experiments provide the possibility of further results \cite{wq}.

\section{Gravitational Tests}
\label{gt}

Sensitivity to 
coefficients $\afb_\mu$, $\cb_\mn$, and $\sbar_\mn$
can be achieved via a wide variety of gravitational experiments \cite{lvpn,lvgap}
including gravimeter experiments \cite{aigrav},
tests of the universality of free fall \cite{bke},
redshift tests \cite{red},
spin-precession tests \cite{jtspin},
experiments with devices 
traditionally used for short-range gravity tests \cite{db},
solar-system tests \cite{solar},
gravity probe B \cite{gpb},
and binary pulsars \cite{joel}.
The key point used in most laboratory tests
is that the gravitational force acquires tiny corrections
both along and perpendicular to the usual free-fall trajectory
near the surface of the Earth due to the above coefficients.
The effective inertial mass of a test body
is also altered in a direction-dependent way
by coefficients $\afb_\mu$ and $\cb_\mn$,
resulting in a nontrivial relation between force and acceleration.
These corrections are time dependent with variations
at the annual and sidereal frequencies,
and
the corrections introduced by
$\afb_\mu$ and $\cb_\mn$
are particle-species dependent.

The properties presented above lead naturally
to a 4-category classification
of 
laboratory tests that use Earth as a source.
Monitoring the gravitational acceleration or force over time
constitutes
free-fall gravimeter tests
or force-comparison gravimeter tests
respectively.
Similarly monitoring the relative acceleration of,
or relative force on,
a pair of test bodies
constitutes a free-fall or a force-comparison Weak Equivalence Principle 
(WEP) test
respectively.
While such tests with ordinary, neutral matter
yield numerous sensitivities to Lorentz violation,
experimentally challenging versions
of the tests highlighted above
performed with  antimatter, charged particles,
and second- and third-generation particles
can yield sensitivities to Lorentz and CPT violation 
that are otherwise difficult or impossible to achieve. 
Gravitational experiments with antihydrogen
\cite{amol2013,aegis,gbar,hbarai,muelcharge,gstate,weax},
charged-particle interferometry \cite{chargeai,muelcharge},
ballistic tests with charged particles \cite{charge},
and signals in muonium free fall \cite{muon}
are considered in Ref.\ \cite{lvgap}.
Here antimatter tests are considered in more detail.

An recent direct measurement of the fall of antihydrogen by the ALPHA collaboration
has placed initial direct limits on differences in the freefall rate
of matter and antimatter \cite{amol2013}.
A number of methods for 
improved measurements are in preparation or have been suggested
including tests using
a Moir\'e accelerometer
\cite{aegis},
trapped antihydrogen 
\cite{gbar},
antihydrogen interferometry 
\cite{hbarai,muelcharge},
gravitational quantum states
\cite{gstate},
and tests in space
\cite{weax}.
As an example of the expected sensitivity
of work in preparation,
the AEGIS experiment
\cite{aegis}
and the GBAR \cite{gbar} experiment both expect
sensitivity at the percent level.

Antimatter-gravity experiments could obtain special sensitivities
to the SME coefficients $\afbw_\mu$ and $\cbw_\mn$.
The key idea is that 
the sign of $\afbw_\mu$ reverses under CPT,
while the sign of $\cbw_\mn$ does not.
Thus,
antimatter experiments
could in principle observe novel behaviors
and could place cleaner constrains
on certain combinations of SME coefficients
than can be obtained with matter.
The analysis of antimatter experiments
is then the same as the treatment of the matter experiments
discussed above
with the only exception being the change in sign
of $\afbw_\mu$ relative its matter counterpart.

In addition to providing a framework 
for the analysis of antimatter gravity experiments
as highlighted above,
the general field-theoretic approach of the SME
illuminates some aspects of attempts 
to place indirect limits on the possibility of unconventional 
antimatter-gravity interactions
that appear in the literature \cite{mntg}.
Toy-model limits
of the SME facilitate the discussion.
One such model is
the isotropic `parachute' model (IPM) \cite{lvgap},
which is similar in design to the IIM discussed above.
The model is constructed
by restricting the classical nonrelativistic Lagrange density of the SME  
in the Sun-centered frame $S$,
to the limit in which only coefficients 
$\afbw_T$ and isotropic $\cbw_{\Si\Xi}$ are nonzero.
The effective classical Lagrangian 
for a test particle T moving in the gravitational field
of a source S
in this limit
can be written in the suggestive form
\beq
L_{\rm IPM} = \half \mt_i v^2 + \fr{\G \mt_g \ms_g}{r}.
\eeq
Here $\mt_i$ is the effective inertial mass of T,
and $\mt_g$ and $\ms_g$ 
are the effective gravitational masses 
of T and S, respectively.
The effective masses are defined in terms of 
the coefficients for Lorentz violation $\afbw_T$, $\cbw_{TT}$ 
and the conventional Lorentz invariant body masses $m^{\rm B}$:
\bea
\nonumber
m^{\rm B}_i &=& 
m^{\rm B} + \sum_w \fracn53 (N^w+N^{\bar{w}}) m^w \cbw_{TT} \\
m^{\rm B}_g &=& 
m^{\rm B} + \sum_w \Big( (N^w+N^{\bar{w}}) m^w \cbw_{TT}
+ 2 \al (N^w-N^{\bar{w}}) \afbw_T \Big),
\label{friedmasses}
\eea
where B is T or S,
$N^w$ and $N^{\bar{w}}$
denote the number of particles and antiparticles of type $w$,
respectively,
and $m^w$ is the mass of a particle of type $w$.

The IPM for electrons, protons, and neutrons,
is then defined by the conditions
\beq
\al \afbw_T = \fracn 13 m^w \cbw_{TT},
\eeq
where $w$ ranges over $e$, $p$, $n$.
This condition
ensure that for a matter body B
the effective inertial and gravitational masses are equal.
That is,
$m^{\rm B}_i = m^{\rm B}_g$,
which implies that no Lorentz-violating effects 
appear in gravitational tests to third post-newtonian order
using ordinary matter;
however,
for an antimatter test body T, 
$m^{\rm T}_i \neq m^{\rm T}_g$
within the IPM.
Thus, 
observable signals may arise in comparisons
of the gravitational responses 
of different types of antimatter
or of the gravitational responses of matter and antimatter.
The following paragraphs
consider the implications for the IPM of some typical arguments against
anomalous antimatter gravity
as well as some new indirect limits recently published.

The question of whether energy remains conserved
when matter and antimatter have different gravitational responses
is one issue that has been raised \cite{pm}.
This point is moot
in the context of the present SME-based discussion
since the energy-momentum tensor is explicitly conserved.
Still, consideration of the following thought experiment is instructive.
Consider a particle and an antiparticle
lowered in a gravitational field,
converted to a photon pair,
raised to the original location,
and reconverted to the original particle-antiparticle pair.
Arguments of this type normally proceed by assuming,
for example,
that the particle-antiparticle pair
gain a particular amount of energy from the gravitational field as they fall,
and
this energy is converted to a pair of photons with no additional change
in gravitational field energy.
The photons are then assumed to couple differently to gravity
than the particle-antiparticle pair,
and hence they lose an amount of energy on the way back to the original height
that differs from the amount gained by the particle-antiparticle pair on the way down,
violating conservation of energy.
To see explicitly how these issues are avoided in the IPM,
first note that in the analysis of Ref.\ \cite{lvgap},
the photons are conventional,
partly via an available coordinate choice.
Next,
note that the CPT-odd coefficient $\afbw_T$
shifts the effective gravitational coupling of the particles
and the antiparticles by equal and opposite amounts relative to the photons.
Thus $\afbw_T$ result in no net difference for the particle-antiparticle combination
and the photons.
To understand
contributions due to the CPT-even coefficient $\cbw_{TT}$,
note that the above argument assumes that no change in potential, or field energy
occurs as the particle-antiparticle pair converts to a photon.
If there is a change in gravitational couplings
this will not typically be the case.
Hence in the case of $\cbw_{TT}$
one finds that energy is conserved when the total energy of the system,
including field energy,
is considered.

Another indirect argument against anomalous antimatter gravity
is based on neutral-meson systems,
which provide natural interferometers
mixing particle and antiparticle states \cite{mlg}. 
Neutral-meson systems have already been used to place tight constraints 
on certain differences of the coefficients $\afbw_\mu$
for $w$ ranging over quark flavors
via flat spacetime considerations
\cite{mesons,akmesons}.
No dominant implications
arise from these constrains
for baryons,
which involve three valence quarks,
or for leptons
in the context of the IPM.
Moreover,
meson tests involve,
valence $s$, $c$, or $b$ quarks,
which are largely irrelevant for protons and neutrons.
This same line of reasoning holds
for gravitational interactions.
Thus the flavor dependence of Lorentz and CPT violation in the SME
implies that the IPM evades restrictions from meson systems. 

As a final popular argument,
consider the attempt to argue against
anomalous antimatter gravity 
based on the large binding energy content of 
baryons, atoms, and bulk matter
\cite{lis2}.
A version of the argument 
relevant for the present discussion of antihydrogen
could be constructed by noting
that the quarks in hydrogen contain only about 10\% of the mass,
with much of the remainder contained 
in the gluon and sea binding.
One might then concluding that 
the gravitational response of the two cannot differ 
by more than about 10\%
since the binding forces are comparable 
for hydrogen and antihydrogen.
Such arguments
implicitly assume
that the gravitational response of a body is determined
by its mass and hence by binding energy.
In the context of the IPM,
the coefficient $\afbw_T$,
leads to a correction to the gravitational force
that is independent of mass,
but can vary with flavor.
Hence the Lorentz-violating modifications to the gravitational responses
are determined primarily by the flavor content
of the valence particles.
One could even consider a scenario in which
the anomalous gravitational effect is associated
purely with the positron,
as would occur in the IPM when $\afbe_T$ is the only nonzero coefficient.
A detailed  consideration of radiative effects
involving $\afbw_T$, $\cbw_{TT}$,
and other SME coefficients for Lorentz violation 
\cite{ck,renorm}
could result in more definite statements along the above lines,
perhaps imposing the IPM condition 
only after renormalization;
however,
the essential points illustrated with the IPM are:
the anomalous gravitational response of a body
can be independent of mass,
can vary with flavor,
and can differ between particles and antiparticles.

Although the IPM provides an example
of a field-theoretic toy model,
which generates an anomalous gravitation response for antimatter
that appears to evade many of the typical indirect limitations,
the model can be limited by a rather different type of
investigation with matter.
Certain experiments with sensitivity to higher powers of velocity \cite{lvgap},
including the recent redshift analysis \cite{red},
considerations of bound kinetic energy \cite{bke},
and double-boost suppression terms,
if analyzed,
in some flat-spacetime tests
can place constraints on the IPM.
Presently the best constraints
are based on bound kinetic energy
and limit the anomalous gravitational response
of antimatter in the IPM to parts in $10^8$ \cite{bke}.
Note however,
that these constraints are rather different from the usual arguments.
It also may be possible to construct models similar to the IPM
based on the recently analyzed higher-order terms in the SME \cite{nonmin},
though this remains an open issue at present.

\section{Summary}

The search for Lorentz and CPT violation
provides the opportunity to probe Planck-scale physics
with existing technology,
and the SME provides a general field-theoretic framework for such investigations.
The comparison of matter and antimatter,
provides a means of conducting such tests
using spectroscopic and gravitational experiments.
Moreover,
a special limit of the SME provides a unique toy model,
the IPM,
for investigating indirect limits on antimatter gravity.


\begin{thebibliography}{}

\bibitem{cpt}
O.W.\ Greenberg,
Phys.\ Rev.\ Lett.\ {\bf 89}, 231602 (2002).

\bibitem{jtrev}
For a recent review, see, J.D.\ Tasson
Rep.\ Prog.\ Phys.\ {\bf 77}, 062901 (2014).

\bibitem{ksp}
V.A.\ Kosteleck\'y and S.\ Samuel,
Phys.\ Rev.\ D {\bf 39}, 683 (1989);
V.A.\ Kosteleck\'y and R.\ Potting,
Nucl.\ Phys.\ B {\bf 359}, 545 (1991).

\bibitem{mav}
N. Mavromatos, 
Hyperfine Interact.\ {\bf 228}, 7 (2014), \thpr.

\bibitem{ck}
D.\ Colladay and V.A.\ Kosteleck\'y,
Phys.\ Rev.\ D {\bf 55}, 6760 (1997);
Phys.\ Rev.\ D {\bf 58}, 116002 (1998).

\bibitem{akgrav}
V.A.\ Kosteleck\'y,
Phys.\ Rev.\ D {\bf 69}, 105009 (2004).

\bibitem{nonmin}
V.A.\ Kosteleck\'y and M.\ Mewes,
Phys.\ Rev.\ D {\bf 80}, 015020 (2009);
Phys.\ Rev.\ D {\bf 85}, 096005 (2012);
Phys.\ Rev.\ D {\bf 85}, 096006 (2013).

\bibitem{lvpn}
Q.G.\ Bailey and V.A.\ Kosteleck\'y,
Phys.\ Rev.\ D {\bf 74}, 045001 (2006).

\bibitem{tables}
{\it Data Tables for Lorentz and CPT Violation,}
2013 edition,
V.A.\ Kosteleck\'y and N.\ Russell,
arXiv:0801.0287v6.

\bibitem{bkr}
R.\ Bluhm, V.A.\ Kosteleck\'y, and N.\ Russell,
Phys.\ Rev.\ D {\bf 57}, 3932 (1998). 

\bibitem{kl}
V.A.\ Kosteleck\'y and C.D.\ Lane,
J.\ Math.\ Phys.\ {\bf 40}, 6245 (1999). 

\bibitem{kr}
V.A.\ Kosteleck\'y and N.\ Russell,
Phys.\ Lett.\ B {\bf 693}, 443 (2010).

\bibitem{akfin}
V.A.\ Kosteleck\'y,
Phys.\ Lett.\ B {\bf 701}, 137 (2011).

\bibitem{lvgap}
V.A.\ Kosteleck\'y and J.D.\ Tasson,
Phys.\ Rev.\ D {\bf 83}, 016013 (2011).

\bibitem{akjt}
V.A.\ Kosteleck\'y and J.D.\ Tasson,
Phys.\ Rev.\ Lett.\ {\bf 102}, 010402 (2009).

\bibitem{clock}
V.A.\ Kosteleck\'y and C.D.\ Lane,
Phys.\ Rev.\ D {\bf 60}, 116010 (1999);
R.\ Bluhm \etal,
Phys.\ Rev.\ Lett.\ {\bf 88}, 090801 (2002). 

\bibitem{hhbar}
R.\ Bluhm, V.A.\ Kosteleck\'y, and N.\ Russell,
Phys. Rev. Lett. 82, 2254 (1999).

\bibitem{bjhyp}
C.\ Malbrunot, 
Hyperfine Interact.\ {\bf 228}, 61 (2014), \thpr;
M.\ Hayden, \thpr.

\bibitem{akifc}
V.A.\ Kosteleck\'y, unpublished (2003).

\bibitem{etrap}
H.\ Dehmelt \etal, 
Phys.\ Rev.\ Lett.\ {\bf 83}, 4694 (1999).

\bibitem{ggtrap}
G.\ Gabrielse \etal, 
Phys.\ Rev.\ Lett.\ {\bf 82}, 3198 (1999);
\thpr. 

\bibitem{wq}
C.C.\ Rodegheri \etal,
Hyperfine Interact. {\bf 194}, 93 (2009).

\bibitem{aigrav}
K.-Y.\ Chung \etal, 
Phys.\ Rev.\ D {\bf 80}, 016002 (2009)

\bibitem{bke}
M. Hohensee, H.\ M\"uller, and R.B. Wiringa
Phys.\ Rev.\ Lett.\ {\bf 111}, 151102 (2013).

\bibitem{red}
H.\ M\"uller \etal, 
Phys.\ Rev.\ Lett.\ {\bf 100}, 031101 (2008);
M.\ Hohensee \etal,
Phys.\ Rev.\ Lett.\ {\bf 106} 151102 (2011);
Phys.\ Rev.\ Lett.\ {\bf 111}, 050401 (2013).

\bibitem{jtspin}
J.D. Tasson 
Phys.\ Rev.\ D {\bf 86}, 124021 (2012).

\bibitem{db}
D.\ Bennett, V.\ Skavysh, and J.\ Long,
in V.A.\ Kosteleck\'y, ed.,
{\it CPT and Lorentz Symmetry V}, 
World Scientific, Singapore, 2010;
H.\ Panjwani, L.\ Carbone, and C.C.\ Speake,
in V.A.\ Kosteleck\'y, ed.,
{\it CPT and Lorentz Symmetry V}, 
World Scientific, Singapore, 2010.

\bibitem{solar}
J.B.R.\ Battat, J.F.\ Chandler, and C.W.\ Stubbs, 
Phys.\ Rev.\ Lett.\ {\bf 99}, 241103 (2007);
A.\ Hees \etal,
arXiv:1301.1658.

\bibitem{gpb}
Q.G.\ Bailey, R.D.\ Everett, and J.M.\ Overduin,
Phys.\ Rev.\ D {\bf 88}, 102001 (2013).

\bibitem{joel}
J.M.\ Weisberg
in V.A.\ Kosteleck\'y, ed.,
{\it CPT and Lorentz Symmetry VI}, 
World Scientific, Singapore, 2014.; 
L.\ Shao and N.\ Wex,
Class.\ Quantum Grav.\ {\bf 30}, 165020 (2013);
L.\ Shao,
Phys.\ Rev.\ Lett.\ {\bf 112}, 111103 (2014).

\bibitem{amol2013}
C.\ Amole \etal,
Nat.\ Commun.\ {\bf 4}, 1785 (2013);
J.\ Fajans, \thpr.

\bibitem{aegis}
AEGIS Collaboration,
A.\ Kellerbauer \etal,
Nucl.\ Instr.\ Meth.\ B {\bf 266}, 351 (2008)
A.\ Knecht \etal, 
Hyperfine Interact.\ {\bf 228}, 121 (2014), \thpr;
J.\ Storey \etal, 
Hyperfine Interact.\ {\bf 228}, 151 (2014), \thpr.

\bibitem{gbar}
P.\ Perez and Y.\ Sacquin, 
Class.\ Quantum Grav.\ {\bf 29},
184008 (2012);
P.\ Indelicato \etal,
Hyperfine Interact.\ {\bf 228}, 141 (2014), \thpr.

\bibitem{hbarai}
AGE Collaboration,
A.D.\ Cronin \etal,
{\it Letter of Intent: 
Antimatter Gravity Experiment (AGE) at Fermilab,} \rm
February 2009;
D.\ Kaplan,
arXiv:1007.4956.

\bibitem{muelcharge}
P.\ Hamilton \etal,
arXiv:1308.1079.

\bibitem{gstate}
A.\ Voronin \etal,
Hyperfine Interact.\ {\bf 228}, 133 (2014), \thpr.

\bibitem{weax}
F.M.\ Huber, E.W.\ Messerschmid, and G.A.\ Smith,
Class.\ Quantum Grav.\ {\bf 18}, 2457 (2001).

\bibitem{chargeai}
B.\ Neyenhuis, D.\ Christensen, and D.S.\ Durfee,
Phys.\ Rev.\ Lett.\ {\bf 99}, 200401 (2007).

\bibitem{charge}
F.S.\ Witteborn and W.M.\ Fairbank,
Phys.\ Rev.\ Lett.\ {\bf 19}, 1049 (1967).

\bibitem{muon}
K.\ Kirch,
arXiv:physics/0702143;
B.\ Lesche,
Gen.\ Rel.\ Grav.\ {\bf 21}, 623 (1989).

\bibitem{mntg}
M.M.\ Nieto and T.\ Goldman,
Phys.\ Rep.\ {\bf 205}, 221 (1991).

\bibitem{pm}
P.\ Morrison,
Am.\ J.\ Phys.\ {\bf 26}, 358 (1958).

\bibitem{mlg}
M.L.\ Good,
Phys.\ Rev.\ {\bf 121}, 311 (1961).

\bibitem{mesons}
KTeV Collaboration,
H.\ Nguyen, 
in V.A.\ Kosteleck\'y, ed.,
{\it CPT and Lorentz Symmetry II}, 
World Scientific, Singapore, 2002
[hep-ex/0112046];
A.\ Di Domenico,
KLOE Collaboration,
J.\ Phys.\ Conf.\ Ser.\ {\bf 171}, 012008 (2009);
FOCUS Collaboration,
J.M.\ Link \etal,
Phys.\ Lett.\ B {\bf 556}, 7 (2003);
BaBar Collaboration,
B.\ Aubert
\etal,
Phys.\ Rev.\ Lett.\ {\bf 100}, 131802 (2008);
hep-ex/0607103.

\bibitem{akmesons}
V.A.\ Kosteleck\'y,
Phys.\ Rev.\ Lett.\ {\bf 80}, 1818 (1998);
Phys.\ Rev.\ D {\bf 61}, 016002 (2000);
Phys.\ Rev.\ D {\bf 64}, 076001 (2001);
V.A.\ Kosteleck\'y and R.\ Van Kooten,
Phys.\ Rev.\ D {\bf 82}, 101702 (2010).

\bibitem{lis2}
L.I.\ Schiff,
Phys.\ Rev.\ Lett.\ {\bf 1}, 254 (1958),
Proc.\ Natl.\ Acad.\ Sci.\ {\bf 45}, 69 (1959).

\bibitem{renorm}
R.\ Jackiw and V.A.\ Kosteleck\'y,
Phys.\ Rev.\ Lett.\ {\bf 82}, 3572 (1999);
M.\ P\'erez-Victoria, JHEP {\bf 0104}, 032 (2001);
V.A.\ Kosteleck\'y, C.D.\ Lane, and A.G.M.\ Pickering,
Phys.\ Rev.\ D {\bf 65}, 056006 (2002);
V.A.\ Kosteleck\'y and A.G.M.\ Pickering,
Phys.\ Rev.\ Lett.\ {\bf 91}, 031801 (2003);
B.\ Altschul, 
Phys.\ Rev.\ D {\bf 69}, 125009 (2004);
Phys.\ Rev.\ D {\bf 70}, 101701 (2004);
B.\ Altschul and V.A.\ Kosteleck\'y, 
Phys.\ Lett.\ B {\bf 628}, 106 (2005); 
H.\ Belich \etal,
Eur.\ Phys.\ J.\ C {\bf 42}, 127 (2005);
T.\ Mariz \etal,
JHEP {\bf 0510}, 019 (2005);
G.\ de Berredo-Peixoto and I.L.\ Shapiro,
Phys.\ Lett.\ B {\bf 642}, 153 (2006);
P.\ Arias \etal,
Phys.\ Rev.\ D {\bf 76}, 025019 (2007);
D.\ Colladay and P.\ McDonald,
Phys.\ Rev.\ D {\bf 75}, 105002 (2007);
Phys.\ Rev.\ D {\bf 77}, 085006 (2008);
Phys.\ Rev.\ D {\bf 79}, 125019 (2009);
M.\ Gomes \etal,
Phys.\ Rev.\ D {\bf 78}, 025029 (2009);
D.\ Anselmi, 
Ann.\ Phys.\ {\bf 324}, 874 (2009);
Ann.\ Phys.\ {\bf 324}, 1058 (2009).


\end{thebibliography}
\end{document}